\documentclass{svmult}

\usepackage{makeidx}         % allows index generation
\usepackage{graphicx}        % standard LaTeX graphics tool
\usepackage{multicol}        % used for the two-column index
\usepackage[bottom]{footmisc}% places footnotes at page bottom
\makeindex             % used for the subject index
\newcommand{\beq}{\begin{equation}}
\newcommand{\beqa}{\begin{eqnarray}}
\newcommand{\eeq}{\end{equation}}
\newcommand{\eeqa}{\end{eqnarray}}
\renewcommand{\a}{\alpha}

\renewcommand{\d}{{\rm d}}
\newcommand{\de}{_{(2)}}

\newcommand\dt[1]{\frad{\d#1}{\d t}}

\newcommand{\frad}[2]{\displaystyle{\displaystyle#1\over\displaystyle#2}}
\newcommand{\g}{\gamma}
\newcommand{\gbar}{\overline{g}}
\newcommand{\infy}{_{(\infty)}}

\newcommand{\meansur}[1]{\langle\!\langle#1\rangle\!\rangle}
\newcommand{\n}{{\bf n}}
\newcommand{\pl}{$\bullet$}
\newcommand{\s}{\sigma}
\newcommand{\one}{_{(1)}}
\newcommand{\vi}{$\circ$}

\newcommand{\N}{{\cal N}}

\begin{document}

\title*{How the rich get richer}

\author{Anita Mehta\inst{1},
A S Majumdar\inst{1}
\and
J M Luck\inst{2}}

\institute{S. N. Bose National Centre for Basic Sciences, Block JD, Salt Lake, Kolkata 700 098, India
\and 
Service de Physique Theorique, CEA Saclay, F 91191 Gif-sur-Yvette, France
 \texttt{anita@bose.res.in, archan@bose.res.in, luck@spht.saclay.cea.fr}}

\maketitle

\begin{abstract}

In our model, $n$ traders interact with each other and with a central
bank; they are taxed on the money they make, some of which is dissipated
away by corruption. A generic feature of our model is that the richest
trader always wins by 'consuming' all the others: another is the existence
of a {\it threshold} wealth, below which all traders go bankrupt.
The two-trader case is examined in detail,
 in the {\it socialist} and {\it capitalist}
limits, which generalise easily to $n>2$.
 In its  mean-field incarnation, our model
exhibits a two-time-scale {\it glassy dynamics},  as well as an
 astonishing {\it universality}.
 When preference is given to local interactions in finite neighbourhoods,
a novel feature emerges: instead of at most one overall winner in the system,
finite numbers of winners emerge, each one the {\it overlord} of a particular region.
The patterns formed  by such winners ({\it metastable states}) are very much
 a consequence of initial
conditions, so that the fate of the marketplace is ruled by its past history;
{\it hysteresis} is thus also manifested.

\end{abstract}

\section{Introduction}
\label{sec:1}
The tools of statistical mechanics~\cite{review} are increasingly being used
to analyse problems of economic relevance \cite{voter}. Our model
below, although originally
 formulated to model the evolution of primordial black holes~\cite{archan, I},
is an interesting illustration of the {\it rich-get-richer} principle
in economics. It is inherently disequilibrating; individual traders interact in such a way that the richest
 trader {\it always} wins.

\section{The model}
\label{sec:2}

In this model,  $n$ traders  
are  linked to each other, as well as to a federal reserve bank;
an individual's money accrues interest at the rate of~$\a>1/2$~\cite{archan}
but is also taxed such that it is depleted at the rate of
 $1/t$, where $t$ is the time.
The interaction strength  $g_{ij}$ between traders
 $i$ and $j$ is a measure of how much of their wealth is invested
in trading; income from trading is also taxed at the rate of $t^{1/2}$.
There is a {\it threshold} term such that the less a trader has, the more he
 loses; additionally the model is {\it non-conservative} such
that some of the wealth disappears forever from the local economy.
These last terms can have different interpretations in a macro-
or a micro-economic context. In the former case (where the traders
could all be citizens of a country linked by a federal bank),
 the threshold term
could represent the plight of the (vanishing) middle classes, while
the non-conservative nature of the model could represent the contribution
 of {\it corruption}
to the economy -  some of the taxed money disappears forever from the region,
to go either to the black economy or to foreign shores. In a more 
micro-economic context (where  traders  linked by a bank are a subset of the
major economy), the interpretation is the reverse: the non-conservative
nature of the model would imply money lost irretrievably 
by taxation (to go to social
benefits from which the traders do not themselves benefit), while the
threshold term could represent the effect of corruption (poorer traders
lose more by graft than richer ones).
Including all these features, we postulate that 
 the wealth
$m_i(t)$ for $i=1,\dots,n$ of each trader  
 evolves as follows~\cite{I}:
\beq
\dt{m_i}=\left(\frac{\a}{t}-\frac{1}{t^{1/2}}\sum_jg_{ij}\dt{m_j}\right)m_i
-\frac{1}{m_i}.
\label{dtm}
\eeq

In the following,we use units of reduced time
$s=\ln\frac{t}{t_0}$ (to renormalise away the effect of
 initial time $t_0$),
 reduced wealth
$x_i=\frac{m_i}{t^{1/2}}$ and reduced square wealth
 $y_i=x_i^2=\frac{m_i^2}{t}$. In these units, we recall
the result for an {\it isolated} trader~\cite{archan}. A trader
 whose initial wealth $y_0$ is  greater than $y_\star$, 
(with $y_\star (t_0)=\left(\frac{2t_0}{2\a-1}\right)$),
is a {\it survivor} who keeps getting richer forever: a trader
with below this threshold wealth goes
 bankrupt and disappears from the marketplace in
 a finite time. The influence of this initial threshold
 $y_\star$ will be seen to persist
throughout this model: in every case we examine, surviving winners
will all be wealthier than this.

\section{A tale of two traders: socialist vs capitalist?}
\label{sec:3}

We examine the two-trader case 
in the {\it socialist} and {\it capitalist} limits.
In the socialist limit, the initial equality of wealth
is maintained forever by symmetry: their common wealth
$x(s)$  obeys:
\beq
x'=\frac{(2\a-1)x^2-2-gx^3}{2x(1+gx)}.
\label{dsx1}
\eeq
This equation is analytically tractable: it has
 fixed points
given by $(2\a-1)x^2-2-gx^3=0$. 
A critical value of the interaction
 strength $g$, 
$g_c=\left(\frac{2(2\a-1)^3}{27}\right)^{1/2}$,
separates two qualitatively different behaviours.
For $g>g_c$, there is
 no fixed point; overly  heavy trading (insufficient saving) causes both 
traders
to go quickly bankrupt, independent of their initial capital.
In the opposite case of sensible trading,
$g<g_c$,
 there are two positive fixed points,
$y_\star^{1/2}<x\one\hbox{ (unstable) }<(3y_\star)^{1/2}<x\de\hbox{ (stable)}.$
If both traders are initial equally poor with
wealth  $x_0<x\one$, this is
 dynamically attracted by  $x=0$ -- the traders go rapidly bankrupt!
 For initially rich traders with $x_0>x\one$, 
their wealth is
 dynamically
 attracted by $x\de$ -- they grow richer forever as
$m(t)\approx x\de t^{1/2}$, a growth rate which
is {\it less} than that for an isolated trader! This
case, where equality and overall prosperity prevail even
though there are no individual winners, could correspond to a (modern)
Marxist vision.

In the {\it capitalist} case, with traders who are initially
unequally wealthy,
any small differences
 always diverge
exponentially  early on:
the details of this transient behaviour can be found in
 \cite{epjb}.
However, the asymptotic behaviour is such that richer trader wins, while the 
poorer one goes bankrupt: 
{\it the survival of the richest is the
 single generic scenario
for two unequally wealthy traders}.
At this point, we are back to the case of an isolated trader referred to in
Section~\ref{sec:3}: he may, depending
on whether his wealth at this point is less or greater than
 $y_\star$,
also go bankrupt or continue to get richer forever.

All of the above generalises easily to 
 any finite number $n\ge2$ of interacting
traders.

\section{Infinitely many traders in a soup - the mean field limit}
\label{sec:4}

We now examine the limit $n\to\infty$: we first explore the
{\it mean field behaviour}
where every trader
is connected to every other by the same dilute
interaction
 $g=\frac{\gbar}{n}$. For fixed $\gbar$, the limit
 $n\to\infty$  leads
to the {\it mean field equations}~\cite{epjb}:
\beq
y'(s)=\g(s)y(s)-2
\label{mfds}
\eeq
When additionally,
$\gbar$ is small (weak trading),
a {\it glassy} dynamics \cite{review} with two-step relaxation is observed. 
In Stage~I, individual traders behave as if they were isolated, so that
 the survivors
are richer than
threshold ($y_\star$), exactly as in the one-trader case 
of Section~\ref{sec:2}.
In  Stage~II, all  traders  interact
 {\it collectively}, and {\it slowly}~\cite{epjb}.
All but the richest trader eventually go bankrupt during this stage.

The model also manifests a striking {\it universality}.
 For example, with an exponential distribution of initial wealth,
 the survival probability  decays asymptotically as
$S(t)\approx\frac{2\a-1}{\gbar}\,\left(C\,\ln\frac{t}{t_0}\right)^{-1/2}$;
additionally,
 the mean wealth of the surviving traders grows as $\meansur{m}_t\approx\left(C\,t\,\ln\frac{t}{t_0}\right)^{1/2}$.
In both cases, $C=\pi$
{\it irrespective} of $\a$, ~$\gbar$ and the parameters of the exponential
distribution. The universality we observe goes further than this, in fact:
it can  be shown~\cite{epjb}  that $C$ only depends on
whether the initial distribution of wealth is bounded or not and on 
(the shape of) the tail of the wealth distribution.

\section{Infinitely many traders with {\it local interactions} - the emergence
of overlords}
\label{sec:5}

Still in the $n\to\infty$ limit, we now introduce local interactions:
traders 
interact preferentially with their
$z=2D$ nearest neighbours on a $D$-dimensional lattice: once again
we look at the limit of weak trading ($g\ll1$).
The dynamics once
again consists of two successive well-separated stages
with fast individual Stage~I dynamics, whose survivors are richer
than threshold, exactly as before (Section~\ref{sec:4}).
The effects of going beyond mean field are only palpable
in Stage II: the effect of local interactions 
lead to a slow
dynamics which
is now very different from the mean-field scenario above.
The survival probability $S(s)$ in fact decays
from its plateau value $S\one$ (number of Stage~I survivors)
 to a non-trivial limiting value $S\infy$;
unlike the mean field result,  a {\it finite} fraction of 
traders now survive forever!

\begin{figure}[htb]
\begin{center}
\includegraphics[angle=90,width=.4\linewidth]{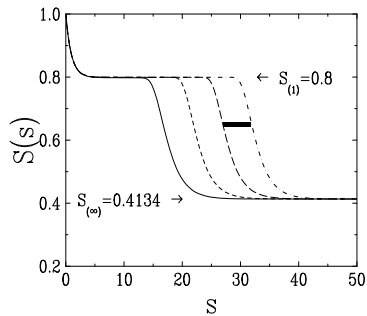}
\caption{\small
Plot of the survival probability $S(s)$ on the chain with $S\one=0.8$ (after
reference~\cite{epjb}.
Left to right:
Full line: $g=10^{-3}$.
Dashed line: $g=10^{-4}$.
Long-dashed line: $g=10^{-5}$.
Dash-dotted line: $g=10^{-6}$.}
\label{figc}
\end{center}
\end{figure}

Figure~\ref{figc}
illustrates this two-step relaxation in the decay of the
 survival probability $S(s)$.
 While the (non-interacting) decay to the plateau
at $S\one=0.8$ is (rightly) independent of  $g$,
the Stage~II relaxation shows {\it ageing}; the weaker
the interaction, the longer the system takes to reach the
 (non-trivial) limit survival probability
$S\infy\approx0.4134$. 

At the end of Stage~II,
the system is left in a non-trivial {\it attractor},
which consists of a pattern where each surviving trader is {\it isolated},
an 
{\it overlord} who keeps getting richer forever.
We call these attractors {\it metastable states},
since they form valleys in the existing random energy landscape;
 the particular metastable state chosen
by the system 
(corresponding to a particular choice of pattern) is the one which
can most easily be reached in this landscape\cite{review}.
The number~$\N$ of these states
generically grows exponentially with the system size (number of sites) $N$ as
$\N\sim\exp(N\Sigma)$
with $\Sigma$  the configurational entropy
or {\it complexity}.
The limit survival probability $S\infy$ (Figure~\ref{figc})
is just the density of a typical attractor,
i.e., the fraction of the initial clusters which survive forever.

We now examine in some more detail the fate of
a set of $k\ge1$ surviving  traders: this depends on $k$ as follows.
\begin{itemize}
\item[$\star$] $k=1$:
If there is only one trader, he
 survives forever, trading with the reserve and getting richer.
\item[$\star$] $k=2$:
If  a pair of neighbouring traders
(represented as~\pl\pl) survive Stage~I,
the poorer dies out, while the richer is an overlord, leading to~{\pl\vi} 
or~{\vi\pl}.
\item[$\star$] $k\ge3$:
If three or more traders survive Stage~I, they may have more than
one fate.
Consider for instance (\pl\pl\pl):
if the middle trader goes bankrupt  first (\pl\vi\pl),
the two end ones are isolated, and both will become overlords.
If on the other hand the trader at the 'end' first goes
bankrupt (e.g.~\pl\pl\vi),
 only the richer among them will become an overlord (e.g.~\pl\vi\vi).
The pattern of these immortal overlords,
and even their number, therefore {\it cannot} be predicted a priori.
\end{itemize}

Finally, we present some of the observed patterns.
If $S\infy=1/2$ on, say, a square lattice,
(i.e. the highest density of surviving traders is reached),
there are only two possible `ground-state' configurations of the system;
 the two possible
patterns of immortal overlords are each perfect checkerboards of one
of two possible parities.
This allows for an interesting possibility: we can  define
a checkerboard index for each site,
which classifies it
 according to its {\it parity}~\cite{epjb}.

\begin{figure}[htb]
\begin{center}
\includegraphics[angle=0,width=.25\linewidth]{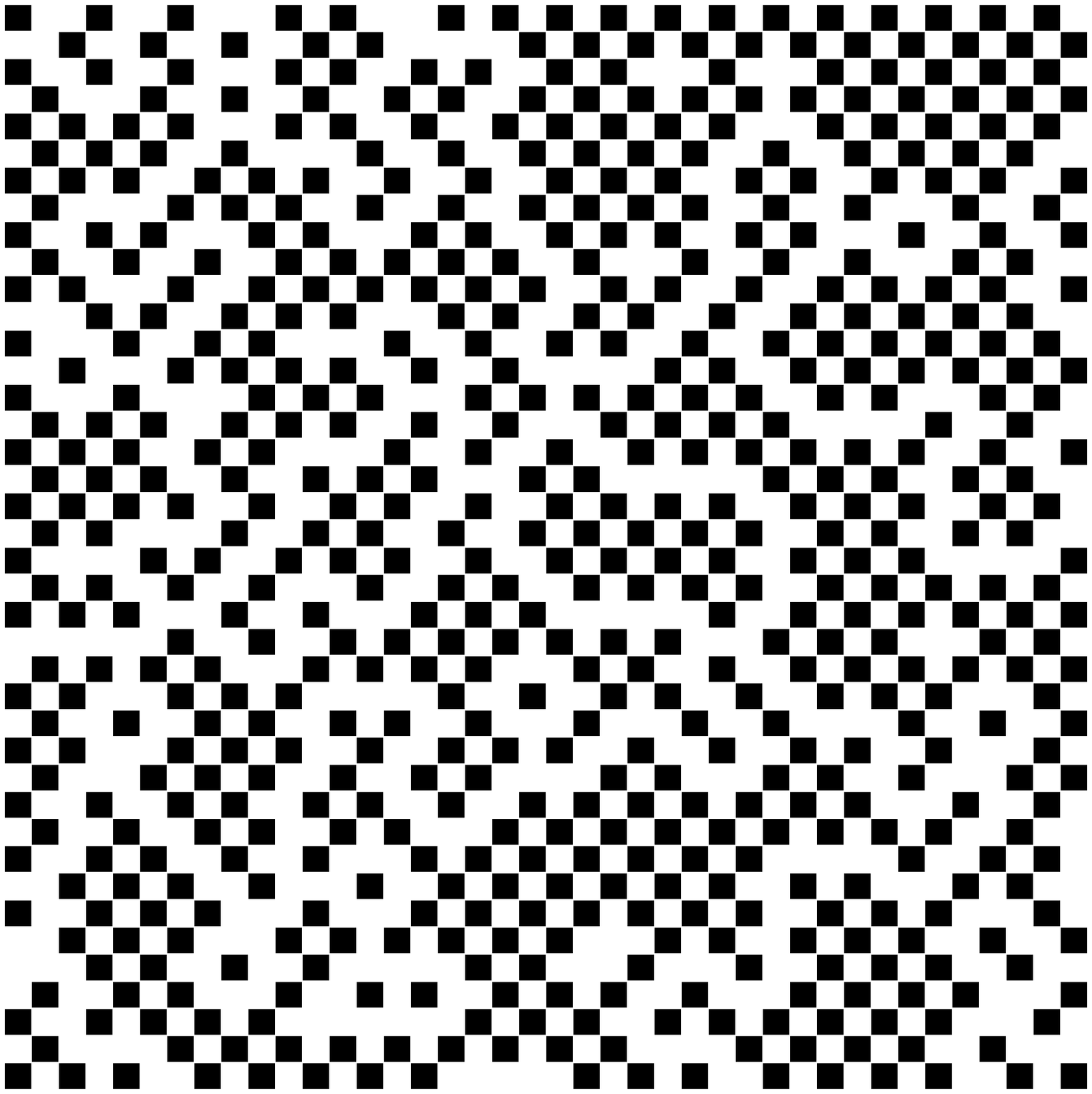}
{\hskip 1mm}
\includegraphics[angle=0,width=.25\linewidth]{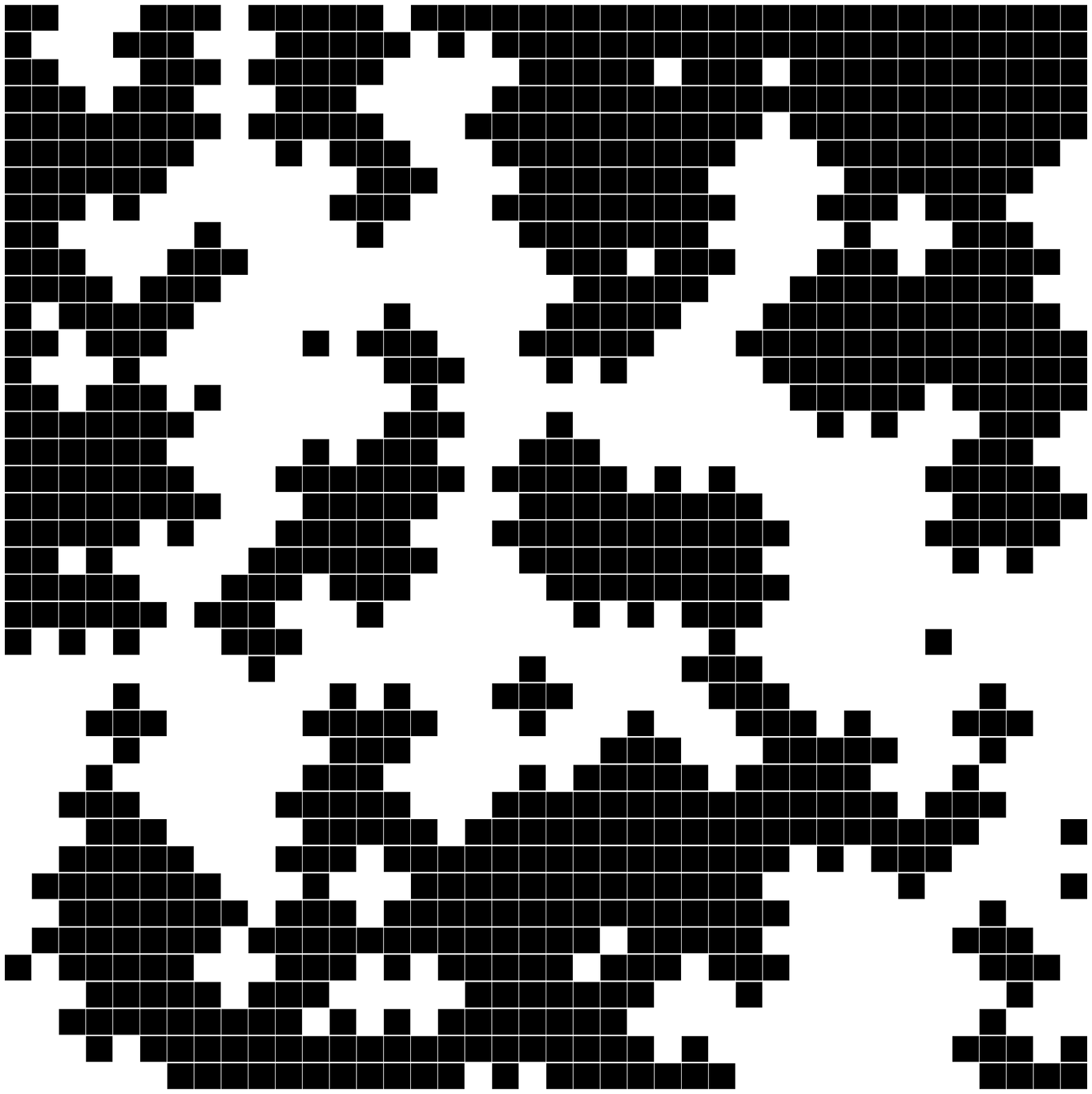}
\caption{\small Two complementary representations
of a typical pattern of surviving clusters on a $40^2$ sample of
the square lattice,
with $S\one=0.9$, so that $S\infy\approx0.371$ (after
reference~\cite{epjb}.
{\bf Left}: Map of the survival index.
Black  squares represent overlords for which $\s_\n=1$,
while white squares represent bankrupt sites for which
$\s_\n=0$.
{\bf Right}: Map of the checkerboard index.
Black squares represent positive, while white squares represent
negative, parity}
\label{fige}
\end{center}
\end{figure}

Figure~\ref{fige} shows a map of the survival index
and of the checkerboard index
for the same attractor for a particular sample of the square lattice.
The {\it local} checkerboard structure,
with random frozen-in defects between patterns of different parities
is of course entirely inherited from the initial conditions.
The overlords in the left-hand part of the figure
are surrounded by rivulets of poverty ; in the right-hand figure, the deviation
from a perfect checkerboard structure (all black or all white)
is made clearer. Neighbouring
sites are fully anticorrelated,
because each overlord is surrounded by paupers:
however, at least close to the limit $S\infy=1/2$,
overlords are very likely to have  next-nearest neighbours 
who are likewise overlords.
The detailed examination of survival and mass correlation functions
made in a longer paper \cite{epjb} confirms these expectations.

To conclude, we have presented a model where traders interact
through a reserve; we are able to model the effects of corruption
and taxation via the non-conservative, threshold nature of our model.
These could have different implications for micro- and macroeconomic
situations.
Our main results are that, in the presence
of global interactions, typically
 only the wealthiest trader survives (provided he was born
sufficiently rich); however, if traders interact locally,
finite numbers of local overlords emerge by creating zones of poverty around
them.

%%%%%%%%%%%%%%%%%%%%%%%%%%%%%%%%%%%%%%%%%%%%%%%%%%%%%%%%%%%%%%%%%%%%%%

\printindex
\end{document}